Mingming Nie, Yijun Xie and Shu-Wei Huang*

# Deterministic generation of parametrically driven dissipative Kerr soliton

**Abstract:** We theoretically study the nature of parametrically driven dissipative Kerr soliton (PD-DKS) in a doubly resonant degenerate micro-optical parametric oscillator (DR-DμOPO) with the cooperation of $\chi^{(2)}$ and $\chi^{(3)}$ nonlinearities. Lifting the assumption of close-to-zero group velocity mismatch (GVM) that requires extensive dispersion engineering, we show that there is a threshold GVM above which single PD-DKS in DR-DμOPO can be generated deterministically. We find that the exact PD-DKS generation dynamics can be divided into two distinctive regimes depending on the phase matching condition. In both regimes, the perturbative effective third-order nonlinearity resulting from the cascaded quadratic process is responsible for the soliton annihilation and the deterministic single PD-DKS generation. We also develop the experimental design guidelines for accessing such deterministic single PD-DKS state. The working principle can be applied to different material platforms as a competitive ultrashort pulse and broadband frequency comb source architecture at the mid-infrared spectral range.

**Keywords:** frequency combs; nonlinear dynamics; optical parametric oscillators; optical solitons; second-order nonlinear optical processes; self-phase locking.

## 1 Introduction

Recent years have witnessed increasing interest in optical frequency comb (OFC) generation in high-Q quadratic nonlinear cavities, benefiting from large $\chi^{(2)}$ quadratic nonlinearity-induced low threshold, high efficiency, and intrinsic frequency conversion. Either intracavity second-harmonic generation [1–6] or optical parametric oscillator (OPO) [7–14] can be utilized for the generation of these quadratic OFCs. In particular, OPO has been demonstrated as a versatile and competitive OFC [2, 15, 16] in otherwise difficult-to-access spectral ranges including the mid-infrared molecular fingerprinting region [17]. Moreover, intriguing dissipative quadratic soliton dynamics in continuous-wave (cw) pumped micro-OPOs (μOPOs) is recently studied and utilized to enhance the performances of ultrafast OPOs [9–14]. These prior μOPO studies are however limited to the operation regime where pump-signal GVM is close to zero and $\chi^{(3)}$ nonlinearity is negligible [12–14].

On the other hand, dissipative Kerr soliton (DKS) formation in fiber-feedback OPO has been recently demonstrated by incorporating the $\chi^{(3)}$ nonlinearity of a meter-long single-mode fiber to balance the cavity dispersion and the $\chi^{(2)}$ parametric gain of a millimeter-long periodically poled lithium niobate (PPLN) to compensate for the cavity loss [18]. This unique example demonstrates how the combination of $\chi^{(2)}$ and $\chi^{(3)}$ nonlinearities in a singly resonant OPO can be utilized to facilitate formation of DKS at signal field that enhances its stability and bandwidth. However, achieving such a clear separation between the $\chi^{(2)}$ and $\chi^{(3)}$ nonlinearities in chip-scale μOPOs is very challenging if not impossible. Furthermore, most chip-scale μOPOs are doubly resonant where both signal and pump are resonant with the cavity and thus the effect of resonant pump must also be considered in the DKS formation dynamics.

In this letter, we theoretically study the nature of DKS at signal field in a cw driven doubly resonant degenerate μOPO (DR-DμOPO) with the cooperation of $\chi^{(2)}$ and $\chi^{(3)}$ nonlinearities. Lifting the assumption of close-to-zero GVM, we show for the first time that there is a threshold

*Corresponding author: Shu-Wei Huang, Department of Electrical, Computer & Energy Engineering, University of Colorado Boulder, Boulder, CO 80309, USA, E-mail: Shuwei.Huang@colorado.edu. https://orcid.org/0000-0002-0237-7317
**Mingming Nie and Yijun Xie,** Department of Electrical, Computer & Energy Engineering, University of Colorado Boulder, Boulder, CO 80309, USA, E-mail: Mingming.Nie@colorado.edu (M. Nie), Yijun.Xie@colorado.edu (Y. Xie)

GVM above which single parametrically driven DKS (PD-DKS) in DR-DµOPO can be generated deterministically. In stark contrast to the previously reported dissipative quadratic soliton [12], deterministic single PD-DKS can stably exist over a wide range of GVM and thus greatly alleviate the need for extensive dispersion engineering. In addition, by simplifying the coupled-wave equation into a single mean-field equation, we clearly identify the different roles of $\chi^{(2)}$ and $\chi^{(3)}$ nonlinearities in the signal PD-DKS formation. With above-threshold GVM, material Kerr nonlinearity (MKN) will dominate the properties of PD-DKS while effective third-order nonlinearity becomes soliton perturbation that is responsible for the soliton annihilation and the deterministic single PD-DKS generation. The exact PD-DKS generation dynamics can be divided into two distinctive regimes depending on the phase matching condition. With large phase mismatch, deterministic single PD-DKS can be obtained with a lower GVM threshold but at the cost of higher pump power. Our theoretical analysis provides the basis for comprehensive understanding of deterministic single PD-DKS generation recently observed in an aluminum nitride (AlN) DR-DµOPO [19]. Finally, we have developed the experimental design guidelines for accessing such deterministic single PD-DKS state, considering various parameters such as GVM, effective second-order nonlinear coefficient and phase mismatch.

## 2 Theoretical analysis and numerical results

The field evolution of a cw-pumped DR-DµOPO with both $\chi^{(2)}$ and $\chi^{(3)}$ nonlinearities (Figure 1) obeys the coupled equations in the retarded time frame of the signal [4]:

$$\frac{\partial A}{\partial z} = \left[ -\frac{\alpha_{c1}}{2} - i\frac{k_1''}{2}\frac{\partial^2}{\partial \tau^2} \right] A + i\kappa B A^* e^{-i\Delta k z} + i(\gamma_1 |A|^2 + 2\gamma_{12}|B|^2)A, \tag{1a}$$

$$\frac{\partial B}{\partial z} = \left[ -\frac{\alpha_{c2}}{2} - \Delta k'\frac{\partial}{\partial \tau} - i\frac{k_2''}{2}\frac{\partial^2}{\partial \tau^2} \right] B + i\kappa A^2 e^{i\Delta k z} + i(\gamma_2 |B|^2 + 2\gamma_{21}|A|^2)B, \tag{1b}$$

and the boundary conditions:

$$A_{m+1}(0, \tau) = \sqrt{1-\theta_1} A_m(L, \tau)e^{-i\delta_1}, \tag{2a}$$

$$B_{m+1}(0, \tau) = \sqrt{1-\theta_2} B_m(L, \tau)e^{-i\delta_2} + \sqrt{\theta_2} B_{in}, \tag{2b}$$

where $A$ is the signal field envelope, $B$ is the pump field envelope, $B_{in}$ is the cw pump, $\alpha_{c1,2}$ are the propagation losses per unit length, $\Delta k'$ is the GVM, and $k_{1,2}''$ are the group velocity dispersion (GVD) coefficients. Without loss of generality, normal GVD at the pump wavelength and anomalous GVD at the signal wavelength are chosen in this work. Higher-order dispersion and nonlinearity are neglected for simplicity. $\kappa = \sqrt{2}\omega_0 d_{\text{eff}}/(A_{\text{eff}}\sqrt{c^3 n_1^2 n_2 \epsilon_0})$ is the normalized second-order nonlinear coupling coefficient, where $d_{\text{eff}}$ is the effective second-order nonlinear coefficient, $A_{\text{eff}}$ is the effective mode area, $c$ is the speed of light, $\epsilon_0$ is the vacuum permittivity, and $n_{1,2}$ are the refractive indices. $\Delta k$ is the wave-vector mismatch, $\gamma_{1,2}$ are self-phase modulation (SPM) coefficients and $\gamma_{12}$ and $\gamma_{21}$ are cross-phase modulation (XPM) coefficients. $L$ is the nonlinear medium length, $\theta_{1,2}$ are the coupler transmission coefficients and $\delta_{1,2}$ are the signal-resonance and pump-resonance phase detuning [4].

By solving the coupled-wave equations via the standard split-step Fourier method (see Section S1, Supplementary material), Gaussian-pulse-seeded generation of DKS pulse profile and optical spectrum at signal field are obtained as shown in Figure 2 (red solid lines). Due to the large GVM, the pump is only quasi-cw without any clear pulse profile (red dashed lines in Figure 2). Importantly, DKS at signal field can only exist when MKNs (SPM and XPM terms) in Eq. (1) are included in the numerical simulation, revealing the key role of MKN in the PD-DKS generation (see Section S2, Supplementary material). We further find that the balance between signal SPM and signal GVD dominates the properties of PD-DKS (blue solid lines in Figure 2). It can be understood as the large GVM renders XPM ineffective and SPM by the weak quasi-cw pump is negligible. Our numerical simulation confirms that the PD-DKS is parametrically driven through the $\chi^{(2)}$ OPO process while its anomalous GVD is balanced by the MKN $\chi^{(3)}$ SPM process. The exact PD-DKS generation dynamics can be divided into two distinctive regimes depending on the phase matching condition.

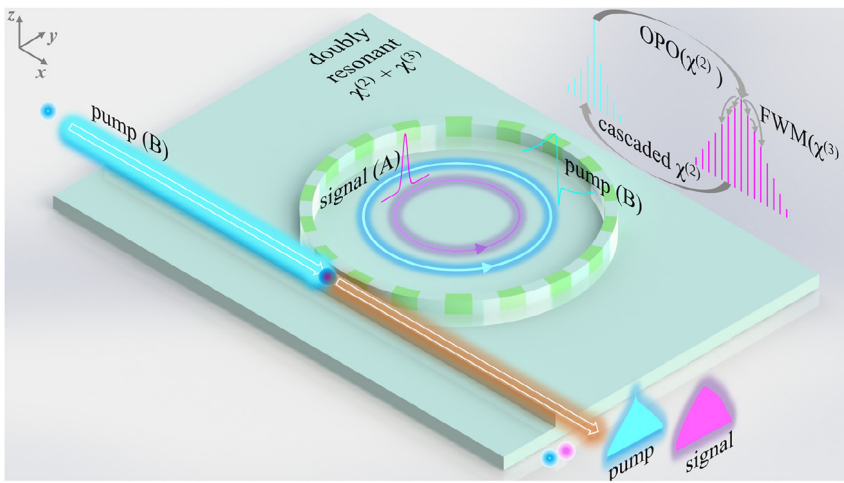

**Figure 1:** Schematic of the deterministic single PD-DKS generation in a DR-DµOPO with both quadratic and cubic nonlinearities. FWM: four-wave mixing.

## 2.1 Deterministic single PD-DKS with perfect phase matching

With near perfect phase matching condition, Eqs. (1) and (2) can be simplified into a single mean-field equation for the signal field, by only considering signal SPM effect (see Section S3, Supplementary material) under the mean-field and good cavity approximations:

$$t_R \frac{\partial A}{\partial t} = \left( -\alpha_1 - i\delta_1 - i\frac{k_1'' L}{2} \frac{\partial^2}{\partial \tau^2} \right) A + i\gamma_1 L |A|^2 A + i\mu A^*$$
$$- [\kappa L \operatorname{sinc}(\xi)]^2 \left[ A^2 \otimes J(\tau) \right] A^*. \quad (3)$$

where $t$ is the "slow time" that describes the envelope evolution over successive round-trips, $t_R$ is the signal roundtrip time, $\tau$ is the "fast time" that depicts the temporal profiles in the retarded time frame, $\alpha_{1,2}$ are the total linear cavity losses. $\mu = \kappa L e^{i(\psi-\xi)} \operatorname{sinc}(\xi) \sqrt{\theta_2} B_{in} / \sqrt{\delta_2^2 + \alpha_2^2}$, is the phase-sensitive parametric pump driving term. Here $\psi = -\arctan(\delta_2/\alpha_2)$ is the phase offset between the cw pump field $B_{in}$ and the signal field $A$, $\xi = \Delta k L / 2$ is the wave-vector mismatch parameter. Equation (3) is the parametrically driven nonlinear Schrödinger equation (PDNLSE) [12–14, 20] with a perturbation term representing effective third-order nonlinearity $J(\tau) = F^{-1}[\hat{J}(\Omega)]$, where $\hat{J}(\Omega) = (\alpha_2 + i\delta_2 - i\Delta k' L\Omega - ik_2'' L\Omega^2/2)^{-1}$ and $\Omega$ is the offset angular frequency with respect to the signal resonance frequency. As shown in Figure 2 with magenta lines, the validity of PDNLSE is verified and it provides a computationally efficient way to study the deterministic PD-DKS generation in this section.

Figure 3(a) and (b) shows the histogram of 100 independent intra-cavity average power traces with and without the last term in Eq. (3), respectively. The pump phase detuning $\delta_2$ ($\delta_2 = 2\delta_1$, Section S4, Supplementary material) is tuned linearly from blue to red side in 180 ns and held constant for another 70 ns to stabilize the PD-DKS generation. Each simulation starts with reinitialized noises to make sure there is no correlation between consecutive runs. Deterministic single PD-DKS formation is observed in Figure 3(a), with each scan converging into the same intra-cavity power and single pulse shape (inset). We emphasize that this deterministic single PD-DKS generation is independent of the pump frequency scanning speed (see Section S5, Supplementary material). In contrast, Figure 3(b) shows that the soliton number is random, exhibiting multiple solitons state or cw state. The results show that the effective third-order nonlinearity, resulting from the cascaded quadratic process, is the main reason for the deterministic single PD-DKS generation.

According to Eq. (3), the perturbation strength of the last term is proportional to the coefficient $[\kappa L \operatorname{sinc}(\xi)]^2$ and $J(\tau)$. We will first study the effect of the latter with a given $d_{eff}$ of 4 pm/V at perfect phase matching condition $\xi = 0$. Similar to our previous study [12], we divide $\hat{J}(\Omega) = X(\Omega) - iY(\Omega)$ into the real and imaginary parts to examine their effect independently, where $X(\Omega)$ and $Y(\Omega)$ resemble the dispersive two photon absorption (TPA) and the dispersive effective Kerr nonlinearity (EKN) respectively. Figure 3(c) and (d) plot the histogram of 100 overlaid intra-cavity average power traces during the frequency scanning process with only EKN and TPA effect, respectively. Deterministic single PD-DKS generation is observed in both simulations, indicating the contribution of both effects. On the other hand, EKN has a more profound effect on the PD-DKS peak power and pulse duration (see the insets of Figure 3), due to its direct impact on the phase detuning, while TPA mainly increases the pump threshold because of the elevated loss. The influence of both effects on the pulse is also investigated with a test Gaussian pulse (see Section S6, Supplementary material) and can be concluded as: (i) a subpulse appears right next to the main pulse; (ii) the intensity of the subpulse can be adjusted by GVM: smaller GVM results in larger subpulse.

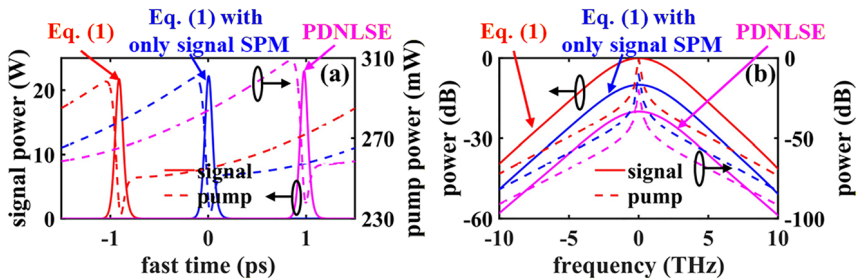

**Figure 2:** Pulse profiles (a) and optical spectra (b) for signal field (solid lines) and pump field (dashed lines), simulated with coupled-wave equations (red lines), coupled-wave equations with only signal SPM (blue lines), and PDNLSE (magenta lines). The pump of PDNLSE is calculated with Eq. (S8). The optical spectra are vertically shifted to emphasize the spectral shapes.

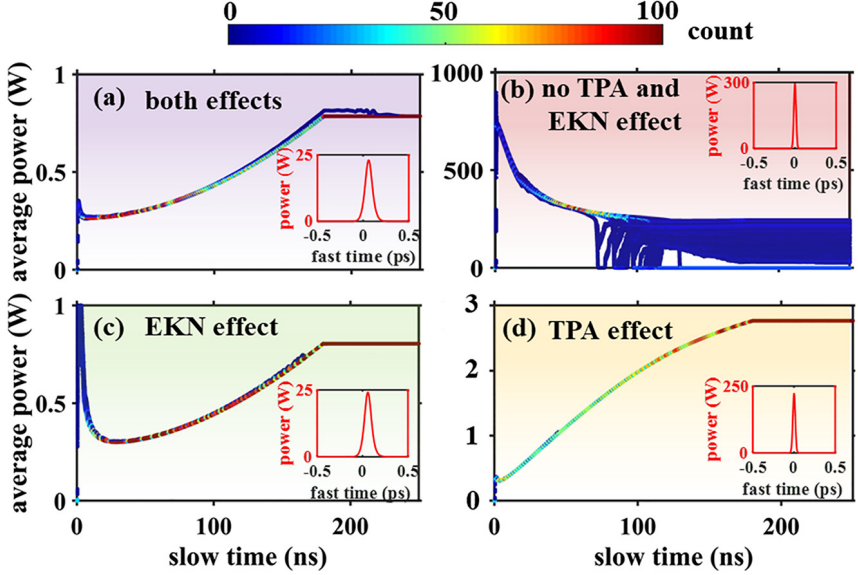

**Figure 3:** Histogram of 100 overlaid intra-cavity average power traces via pump frequency scanning. $|B_{in}|^2$ = 30 mW. (a) With both TPA and EKN effect; (b) without no TPA and EKN effect; (c) with only EKN effect; (d) with only TPA effect. The insets are the pulse profile of stable soliton. In the inset of (b), one of the multiple solitons is chosen.

As shown in Figure 4(a) and (b), with large GVM $X(\Omega)$ and $Y(\Omega)$ are both narrowband with maximum values at the center frequency. Therefore, multiple solitons will experience long range interaction due to narrowband perturbation [21]. According to Eq. (3), this perturbation from TPA or EKN effect can be viewed qualitatively as amplitude or phase modulation to the pump field ($i\mu A^*$), which is a common method for deterministic DKS generation in $\chi^{(3)}$ nonlinear cavities [22–24]. Furthermore, the last term in Eq. (3) breaks down the phase symmetry $A \to -A$, which means solitons with opposite phase can no longer exist. Simulations show that single soliton with opposite phase will disappear with TPA or automatically adjust its phase and evolve into a soliton with EKN.

Figure 4(c) and (d) show how multiple solitons from Figure 3(b) evolve under the influence of TPA and EKN, respectively. Similar to the avoided mode crossings induced Cherenkov radiation [25], TPA and EKN lead to dispersive waves and destabilize the solitons through long range interaction. Multiple solitons interact with each other, experience extra loss from the dispersive waves and finally only single soliton survives. Breathing behaviors are observed during the soliton interaction process, which is common due to the energy exchange between multiple solitons [25]. In addition, soliton interaction is more sensitive to EKN rather than TPA, similar to the more effective pump phase modulation method for conventional DKS generation. In fact, during the pump frequency scanning process in Figure 3(c) and (d), single soliton usually arises from the highest intensity peak in the background (see Section S5, Supplementary material) instead of multiple solitons, resulting in no evident soliton steps for soliton annihilation.

Bandwidth of $X(\Omega)$ and $Y(\Omega)$ increases with the reduction of GVM, thus causing stronger soliton perturbation. With $d_{eff}$ = 4 pm/V, a GVM larger than 380 fs/mm is required to keep the soliton perturbation manageable so the PD-DKS can still sustain itself. With below-threshold GVM of 300 fs/mm, the PD-DKS breaks up into subpulses and eventually evolves into a cw solution [see Figure 4(e)]. Importantly, the GVM threshold increases as a function of $d_{eff}$ [Figure 4(f)] because larger $d_{eff}$ means stronger soliton perturbation [Eq. (3)] and in turn requires larger GVM to alleviate the perturbative effect and stabilize the PD-DKS.

## 2.2 Deterministic single PD-DKS with large phase mismatch

In the case of large phase mismatch, the single mean field equation Eq. (3) fails to describe the system dynamics as the coherence length is much smaller than the cavity length. The integration along the cavity length from Eq. (1) to Eq. (S6) (see Section S3, Supplementary material) and the averaging effect of laser fields is invalid due to the strong energy exchange between pump and signal within one roundtrip. The comparative results of Eqs. (1) and (3) in Figure S5 (see Section S7, Supplementary material) indicate that PDNLSE is a good approximation only around the perfect phase matching point. Therefore, we will apply Eq. (1) along with the boundary conditions Eq. (2) to investigate deterministic PD-DKS generation with large phase mismatch in this section.

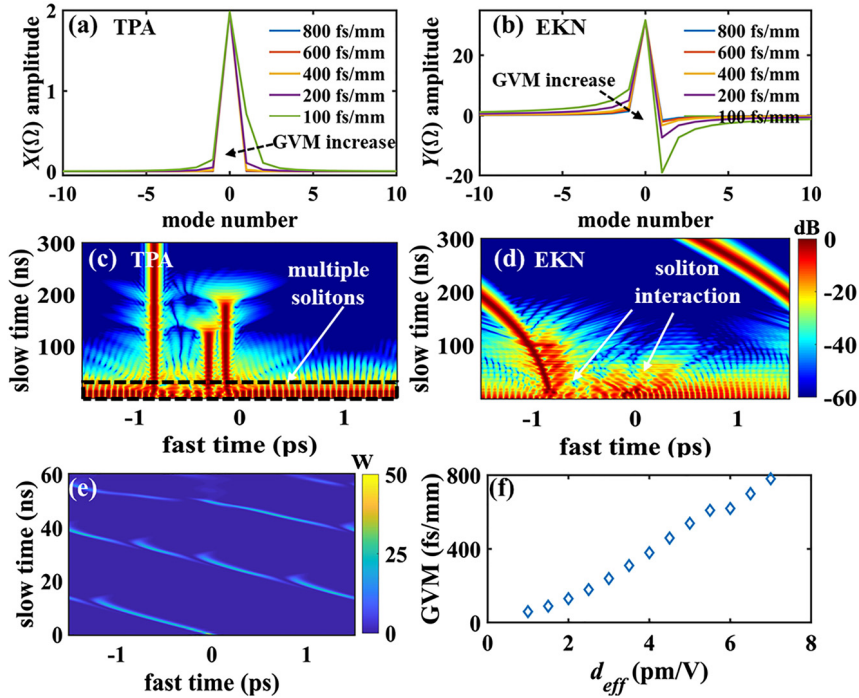

**Figure 4:** The influence of effective TPA X(Ω) (a) and EKN Y(Ω) (b) on the cavity modes. Single soliton evolves from multiple solitons as a result of Figure 2(b) under the effective TPA (c) and EKN (d). Both effects are increasing from zero to the maximum amplitudes. (e) Pulse break-up due to large perturbation from effective third-order nonlinearity with a smaller GVM = 300 fs/mm. (f) GVM threshold versus second-order nonlinear coefficient.

PD-DKS can also be achieved with large phase mismatch, at the cost of larger pump threshold. As shown in Figure 5(a) and (b), the pump is no longer quasi-cw but becomes a Turing pattern with large modulation depth, corresponding to a strong spectral peak at $\Delta\omega = 2\pi/(\Delta k'L)$. Intuitively, Turing rolls is generated through GVM induced MI [2] where the modulation depth increases with $\xi$. As for temporal dynamics, Figure 5(c) and (d) show the evolution of signal and pump field within three successive roundtrips. In the retarded time frame of the signal, the pump Turing roll drifts at a speed of $1/\Delta k'$ in the cavity but remains locked to the PD-DKS after each roundtrip, resulting from the temporal separation of $\Delta k'L$ between adjacent Turing rolls. This can be understood by regarding Turing rolls as potential wells where the PD-DKS just hops across one during each roundtrip. The pulse duration of PD-DKS gets shorter with higher peak power when GVM decreases as Turing rolls become narrower. In addition, the intracavity pump experiences a small temporal shift [see the inset of Figure 5(d)] at the coupler region when it meets with the external cw drive.

Figure 5(e) shows the histogram of 100 overlaid intracavity average power traces with pump frequency tuning. Deterministic single PD-DKS formation is observed in simulations. The determinism still comes from the perturbative effective third-order nonlinearity, which can be understood qualitively with Eq. (3) although it cannot exactly describe the system dynamics. Of note, large phase mismatch reduces the magnitude of effective third-order nonlinearity [12] and in turn lowers the GVM threshold. Comparing Figure 5(f) with Figure 4(f), it is shown the threshold GVM above which single PD-DKS can be deterministically generated is lowered by almost an order of magnitude when a wave-vector mismatch parameter $\xi$ of $0.75\pi$ is introduced.

### 2.3 Experimental guidelines

Based on the previous analysis, moderate perturbation induced by cascaded quadratic process is necessary to achieve deterministic single PD-DKS generation. Stronger perturbation leads to complete soliton destabilization while weaker perturbation results in multiple soliton generation. The interplay between $d_{eff}$, GVM, and $\xi$ uniquely define the evolution dynamics. Strong perturbation from effective third-order nonlinearity can be alleviated by lowering the strength of cascaded quadratic process through the reduction of $d_{eff}$ or the increase of $\xi$ and GVM.

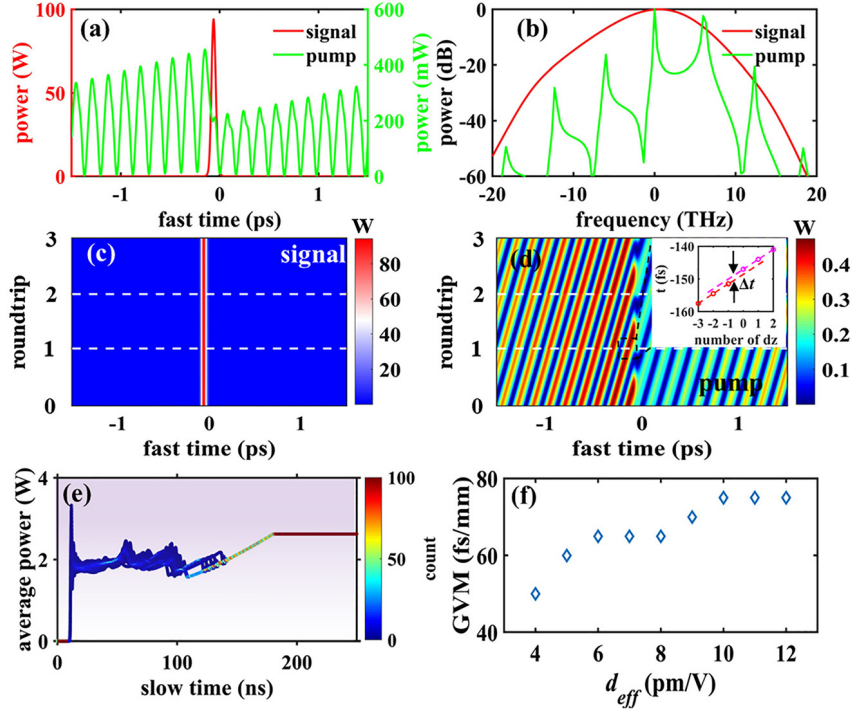

**Figure 5:** Signal soliton with large phase mismatch. $\xi = 0.75\pi$, $|B_{in}|^2 = 100$ mW. (a) Pulse profiles and (b) spectra. Pulse evolution within three successive roundtrips (indicated by white dashed lines) for signal (c) and pump field (d). The inset of (d) shows the temporal positions in time axis corresponding to peak powers before and after meeting with pump. (e) Histogram of 100 overlaid intra-cavity average power traces with same frequency scanning strategy in Figure 2. (f) GVM threshold versus second-order nonlinear coefficient.

To experimentally access the deterministic single PD-DKS state, one has to consider how these three parameters co-determine the system's behavior as well as the experimental restrictions: (i) GVM can be tuned via dispersion engineering [26] but it is ultimately limited by material dispersion; (ii) $d_{eff}$ can be changed by choosing different nonlinear crystals or different crystal axes; (iii) $\xi$ is the most controllable parameter in experiment, through temperature tuning, angle tuning, quasi-phase-matching, and more. Admitting the above-mentioned experimental restrictions, the design rules to experimentally access deterministic PD-DKS can be summarized as: (i) for small $d_{eff}$ such that $\kappa^2 L \ll \gamma_1$, it is better to operate near the perfect phase matching point to lower the pump threshold; however, the GVM threshold is higher in this regime; (ii) for large $d_{eff}$ such that $\kappa^2 L \gg \gamma_1$, it is better to operate with large phase mismatch to lower the GVM threshold; compromise between the GVM threshold and pump threshold should be considered in this regime.

Practically speaking, soliton microcomb generation is always accompanied by a pronounced thermal effect due to the small mode volume. In Section S8 of the Supplementary material, we include the equations of thermally induced phase detuning and phase mismatch and solve them together with the two-wave coupled equation to understand the thermal effect. It is found that the thermal effect neither inhibits the deterministic single PD-DKS generation nor changes the PD-DKS properties. However, due to the shift of phase detuning and phase mismatch resulting from the slow thermal dynamics, the initial phase mismatch, frequency scanning speed, and final phase detuning should be carefully chosen in real experiments.

### 2.4 PD-DKS evolution dynamics in AlN DR-DμOPO

A very recent experiment using an AlN DR-DμOPO has shown deterministic soliton generation [19] and our numerical analysis reveals its PD-DKS nature with the steady-state pulse profile and optical spectrum shown in Figure S11 (Section S9, Supplementary material) and the temporal and spectral evolution dynamics shown in Figure 6. The evolution dynamics are obtained by scanning the pump frequency from the blue to red detuning sides of the cavity resonance and the phase mismatch from $\xi = 0.75\pi$ to $\xi = 0.05\pi$ as a consequence of intracavity temperature change [27, 28] during the pump frequency scan. Excellent agreements with the experimental observation [19] are achieved. In particular, its determinism is the consequence of large MKN, large GVM, and small $d_{eff}$. Our numerical analysis evidently shows several localized pump spectral modulations [dashed lines in Figure 6(b)], which are also observed experimentally [19] but not further examined and well understood.

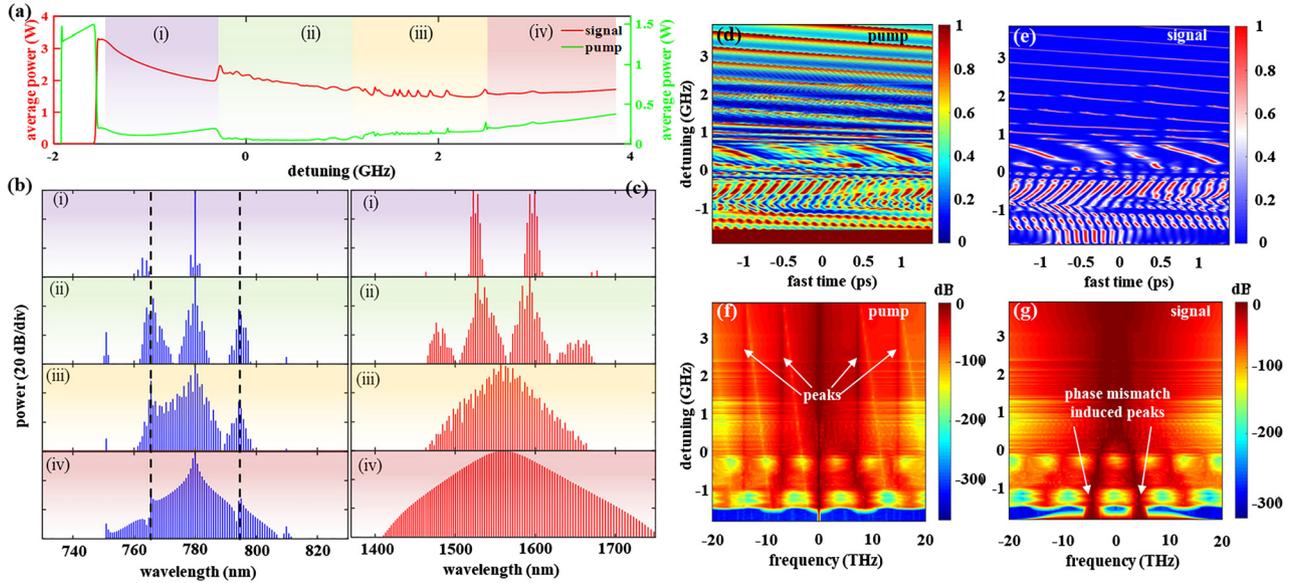

**Figure 6:** Dynamics during the pump frequency scanning process.
(a) Intra-cavity average power evolves with pump frequency detuning. (b), (d), (f): Pump field, (c), (e), (g): Signal field. (b) and (c) show the pump and signal field spectra for the four distinctive states (i)–(iv), respectively. (i): Turing pattern; (ii) and (iii): Chaotic states; (iv) single PD-DKS. The dashed lines in (b) show the unchanged pump spectral peaks. (d) and (e) show the pump and signal field temporal intensity evolution with pump frequency scanning from blue to red side, respectively. (f) and (g) show the pump and signal field spectral intensity evolution with pump frequency scanning from blue to red side, respectively. Intensity is normalized for each roundtrip in (d)–(g).

In time domain, the signal field successively experiences Turing patterns with many rolls [state (i) in Figure 6(b)], chaotic states [state (ii) and (iii)] and finally single pulse evolving into single PD-DKS [state (iv)], as shown in Figure 6(e). In frequency domain, firstly, there are two symmetric signal spectral peaks [Figure 6(g)] resulting from the nondegenerate OPO process due to the large phase mismatch. With the decreasing of phase mismatch, the two peaks move close and then merge with each other at the degenerate frequency [Figure 6(g)]. When the signal is deeply red-detuned, single soliton emerges and results in a broadband spectrum [Figure 6(g)].

For pump spectrum, firstly, there is only a pair of pump spectral peaks in the nondegenerate OPO process [Figure 6(f)]. With the decreasing of phase mismatch, spectral peaks located at other positions arise due to the energy exchange between the $\chi^{(3)}$ nonlinearity assisted signal spectrum broadening and the $\chi^{(2)}$ process, among which the three frequencies (pump frequency $2\omega_0 + \Delta\omega$, signal frequencies $\omega_0$ and $\omega_0 + \Delta\omega$) start to mix with each other and result in pump spectra peaks located at an integer multiple of $2\pi/(\Delta k'L)$. Once the signal spectrum is broad enough, the spectral peaks keep their positions.

## 3 Conclusions

In conclusion, we theoretically study the nature of PD-DKS generation in a cw driven DR-DµOPO with the cooperation of $\chi^{(2)}$ and $\chi^{(3)}$ nonlinearities. We show for the first time that there is a threshold GVM above which single PD-DKS in DR-DµOPO can be generated deterministically. In stark contrast to the previously reported dissipative quadratic soliton [12], deterministic single PD-DKS can stably exist over a wide range of GVM and thus greatly alleviate the need for extensive dispersion engineering. In addition, by simplifying the coupled-wave equation into a single mean-field equation, we clearly identify the different roles of $\chi^{(2)}$ and $\chi^{(3)}$ nonlinearities in the signal PD-DKS formation. With above-threshold GVM, MKN will dominate the properties of PD-DKS while effective third-order nonlinearity becomes soliton perturbation that is responsible for the soliton annihilation and the deterministic single PD-DKS generation. Similar effect has also been observed in the DKS platform where the fundamental-second-harmonic mode coupling promotes the deterministic single DKS generation [29, 30]. The exact PD-DKS generation dynamics can be divided into two distinctive regimes depending on the phase matching condition. In both regimes, the perturbative effective third-order nonlinearity resulting from the cascaded quadratic process is responsible for the soliton annihilation and the deterministic single PD-DKS generation. Moreover, with large phase mismatch, deterministic single PD-DKS can be obtained with reduced GVM threshold but at the cost of higher pump power. To access the deterministic single PD-DKS state, it is thus better to operate near the perfect phase matching point with low $d_{\text{eff}}$ materials while it is beneficial to operate at large phase mismatch when high $d_{\text{eff}}$ materials are available. Benefiting from the low complexity and wavelength down-conversion of PD-DKS, the working principle can be applied to different material platforms as a competitive field-deployable ultrashort pulse and broadband frequency comb source architecture at the mid-infrared molecular fingerprinting spectral range.

**Author contributions:** All the authors have accepted responsibility for the entire content of this submitted manuscript and approved submission.
**Research funding:** This work has been supported by the Office of Naval Research (ONR) under award number N00014-19-1-2251.
**Conflict of interest statement:** The authors declare no conflicts of interest.

**Supplementary Material:** The online version of this article offers supplementary material (https://doi.org/10.1515/nanoph-2020-0642).